\documentclass[twocolumn]{article}
\usepackage{graphicx}
\begin{document}

\title{Adiabatic transport of Cooper pairs in arrays of 
Josephson junctions} 

\author{J.P. Pekola, J.J. Toppari, M. Aunola, and M.T. Savolainen \\~ \\
{\sl Department of Physics, University of Jyv{\"a}skyl{\"a}, P.O. Box 35,
 40351 Jyv{\"a}skyl{\"a}, Finland}\\~ \\
 D.V. Averin \\~ \\ \sl Department of Physics, Harvard University, 
Cambridge, 
MA 02138; and \\ 
Department of Physics and Astronomy,  SUNY at Stony Brook, 
Stony Brook NY 11794 }
\maketitle
{\sl
We have developed a quantitative theory of Cooper pair pumping in 
gated one-dimensional arrays of Josephson junctions. The pumping 
accuracy is limited by quantum tunneling of Cooper pairs out of the 
propagating potential well and by direct supercurrent flow through 
the array. Both corrections decrease exponentially with the number 
$N$ of junctions in the array, but give a serious limitation of 
accuracy for any practical array. The supercurrent at resonant gate 
voltages decreases with $N$ only as $\sin(\varphi/N)/N$, where   
$\varphi$ is the Josephson phase difference across the array.}  

{\small\hfill PACS numbers: 73.23.Hk, 74.50.+r}

When a potential well propagates adiabatically along an electron system 
that is effectively one-dimensional, it carries with it additional 
electron density and, in this way, induces a flow of dc electric current 
through the system. Such a pumping effect is observed in a large variety 
of mesoscopic systems ranging from small metallic tunnel junctions in 
the Coulomb blockade regime \cite{b1,b2,b3}, to semiconductor quantum 
dots \cite{b4}, and to one-dimensional ballistic channels \cite{b5}. 
The propagation of potential well in these systems is arranged either 
directly through the propagation of real acoustoelectric wave 
\cite{b2,b5} or by a few (two or more) phase-shifted 
gate voltages \cite{b1,b3,b4}. Particular interest is attracted to the 
pumping regime when the potential well carries a quantized number $m$ of 
electrons so that the induced current $I$ is related to the frequency 
$f$, with which the well crosses the system, by fundamental relation 
$I=mef$. The quantization of $m$ is caused by the existence of the 
energy gap in the spectrum of the underlying electron system, which 
makes it possible for the potential well to be deep enough for 
precisely $m$ electrons. Such an energy gap can be created either by 
the Coulomb interaction, as, for instance, in the Coulomb blockade 
pumps \cite{b1,b2,b3}, or can be simply due to the discrete nature 
of single-particle states inside the well \cite{b6,b7}. In the case 
of Coulomb blockade pumps, precision of the pumped charge 
quantization is reaching the level sufficient for metrological 
applications \cite{b3}. Different sources of inaccuracy in the pumps 
have been discussed in \cite{a1,a2,a3,a4,b3}. 

Until presently, the pumping effect was studied mostly in normal 
systems, where transport is due to individual electrons. A timely 
motivation for studying Cooper pair transfer comes from quantum 
computation, where pumping can play an important role as an essential 
element of dynamics of quantum logic gates \cite{b8*}. The aim 
of this work is to develop a quantitative theory of pumping of 
Cooper pairs in one-dimensional arrays of superconducting tunnel 
junctions. In particular, we find fundamental corrections to the quantized 
pumping regime and show that they are 
unexpectedly large in arrays with a small number of junctions. These 
large quantum corrections can also explain the fact that the first 
(and so far the only) experiment with pumping of Cooper pairs 
\cite{b8} has failed to demonstrate accurate pumping.    

First we derive the general expression for the charge 
transferred through an array of $N$ superconducting tunnel 
junctions in the Coulomb blockade regime by adiabatic pumping of 
Cooper pairs. In the standard model such arrays are 
characterized by two energies, their charging energy $H_{\rm C}$ 
as a system of capacitors, and their energy associated with tunneling 
\cite{b9}. We assume that the characteristic energy $E_{\rm C}$ of a 
Cooper pair in the array and the temperature ($k_{\rm B}T$) are both 
much lower 
than the superconducting energy gap of the electrodes. The first 
one replaces the usual condition in the
normal Coulomb blockade where the junction resistances should be 
larger than the quantum resistance. With these conditions fulfilled 
quasiparticle tunneling is exponentially suppressed, while 
the tunneling energy of Cooper pairs in junction $i$ reduces to a 
constant tunneling amplitude $E_{{\rm J}i}/2$. ($E_{{\rm J}i}$ is also 
called Josephson coupling energy.) In this work, the bias voltage is 
set to be zero, and thus a constant Josephson phase difference $\varphi$ 
is fixed across the array. We can then treat the two external electrodes 
of the array as one, so that effectively the array forms a loop and 
$\varphi$ plays the role of external flux threading it. Then, the 
Hamiltonian of the N-pump is: 
\begin{equation}
H\! =\!H_{\rm C}(n\!-\!q)-\!\!\! \sum_{k=1}^N \!\frac{E_{{\rm J}k}}{2} 
\left( 
|n\rangle \langle n+\delta_k| e^{i\varphi/N}\!\! + \!h.c. \right) . 
\label{1} \end{equation}
Here $n\equiv \{n_1,n_2,\ldots,n_{N-1} \}$ and $q\equiv \{q_1,q_2,
\ldots,q_{N-1} \}$ represent, respectively, the number $n_i$ of 
Cooper pairs on each island of the array, and 
$q_i$, normalized by $2e$, the charge injected to each island by the gate 
voltage $V_{{\rm g}i}$ (Fig.\ 1). The term $\delta_k$ describes the 
change of $n$ due to tunneling 
of one Cooper pair in the $k$th junction. We will also need the 
operator of the  current in the $k$th junction: 
\begin{equation}
I_k= \frac{i eE_{{\rm J}k}}{2\hbar} \left(|n\rangle \langle n+\delta_k| 
e^{i\varphi/N} - h.c. \right) \, . 
\label{2} \end{equation}
\begin{figure}[ht]
\center
\includegraphics*[width=70mm]{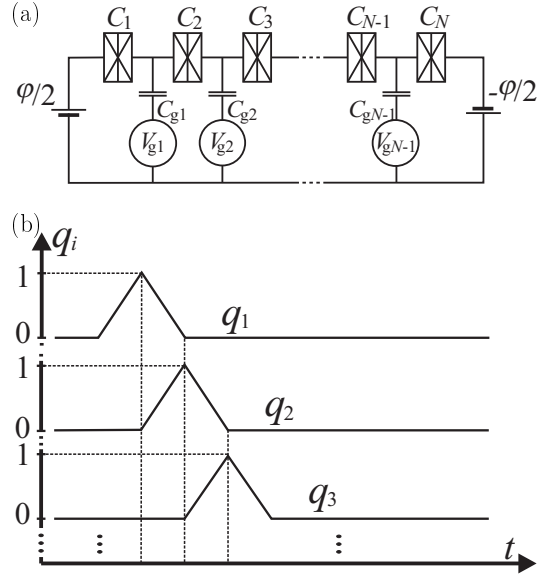}
\caption{\small {\bf (a)} A schematic drawing of a gated Josephson array 
of $N$ junctions. In pumping Cooper pairs gate voltages $V_{{\rm g}i}$ 
are operated cyclically. 
$C_i$ are the capacitances of the junctions, and $C_{{\rm g}i}$ are gate
capacitances. In a uniform pump $C_i \equiv C$ for all $i=1,2, 
\ldots,N$. {\bf (b)} A train of gate voltages to carry a charge in a pump. 
Here $q_i = - C_{{\rm g}i}V_{{\rm g}i}/2e$.}
\end{figure}

There are two mechanisms of Cooper pair transport in the array. One is 
the direct supercurrent through the whole array, another is pumping, 
the charge transfer in response to adiabatic variation of the 
injected charges $q$. To derive the 
general expression for the total charge $Q$ transferred during 
one pumping period, we introduce the basis of instantaneous 
eigenstates $|m\rangle$ of the array for a given $q$. These are exact 
eigenstates if $q$ is stationary. When 
$q$ varies, there is a correction  $|\delta m\rangle$ to state 
$|m\rangle$ of the form: 
\begin{equation}
|\delta m \rangle = i\hbar \sum_{l \neq m}\frac{ |l\rangle \langle 
l|\nabla_q m \rangle }{\varepsilon_l -\varepsilon_m} \dot{q} \, ,   
\label{3} \end{equation}
where $\varepsilon_{l,m}$ are the eigenenergies of states 
$|l,m\rangle$. Thus the average current 
$\langle I_k\rangle$ in the $k$th junction is different from the 
dc supercurrent $\langle m|I_k|m\rangle$ in the state $|m\rangle$, 
$ \langle I_k \rangle = \langle m|I_k|m\rangle +2\mbox{Re} 
\langle m|I_k|\delta m\rangle$.    
Integrating over the pumping period $\tau$ we obtain the charge $Q$, 
in units of $2e$, 
transported through the array: 
\begin{equation}
Q = \frac{1}{\hbar} \frac{\partial }{ \partial \varphi } 
\int_0^{\tau} dt \varepsilon_m (q(t)) + 2 \mbox{Re} \oint 
\langle m|Q_k|d m \rangle \, .  
\label{4} \end{equation}
Here we used the standard thermodynamic argument that the 
supercurrent in the state $|m\rangle$ can be expressed as 
$(2e/\hbar)\partial \varepsilon_m /\partial \varphi$, and defined  
the operator of the normalized charge transferred in the $k$th 
junction, $Q_k= (1/2e) \int dt I_k (t)$: 
\begin{equation}
\langle m|Q_k|l\rangle = \frac{i\hbar}{2e} 
\frac{\langle m|I_k|l\rangle}{\varepsilon_l -\varepsilon_m} \, . 
\label{5} \end{equation} 
Diagonal matrix elements of $Q_k$ are not defined by eq.\ 
(\ref{5}), but they do not 
contribute to $Q$.  

Few remarks should be made concerning eq.\ (\ref{4}). Since the 
state of the array at the end of the pumping cycle is exactly the 
same as in the beginning, charge conservation implies that $Q$ is 
independent of the index $k$ of the 
junction which is used in calculating $Q$. It is,  
however, a function of the array state $|m\rangle$. We did not 
include energy relaxation in the model and pumping is in principle 
possible even when $|m\rangle$ is an excited state. Below we 
consider only a more typical situation when the array is in 
equilibrium so that at low temperatures $|m\rangle$ can only be a 
ground state. The last remark is that eq.\ (\ref{4}) shows that 
there is a close 
connection between the transferred charge $Q$ and quantum 
mechanical phase of the state $|m\rangle$ accumulated during 
the cycle. Supercurrent contribution to $Q$ is 
directly related to the dynamic part of the phase, 
$\int_0^{\tau} dt \varepsilon_m (q(t))/\hbar$, while the pumped 
charge is associated with the Berry's phase $\xi= i \oint 
\langle m|d m \rangle$ \cite{b10}. 

In the following quantitative analysis, the array is assumed to 
be uniform. The charging energy of the array is then:   
\begin{equation}
H_{\rm C}\! =\!\!\frac{E_{\rm C}}{N} \!\left [ \sum_{k=1}^{N-1}\!\! 
k(N\!\!-\!k) u_k^2\! + \!2\!\!\! 
\sum_{l=2}^{N-1} \sum_{k=1}^{l-1} k(N\!\!-\!l) u_ku_l \!\right ] ,
\label{6} \end{equation} 
where $E_{\rm C}\equiv (2e)^2/2C$, $C$ is the common capacitance of 
each junction in the array, 
and $u_k\equiv n_k-q_k$. In the regime of accurate pumping the main 
contribution to 
$Q$ comes from the second term in (\ref{4}) while the 
supercurrent gives only small corrections limiting the pumping 
accuracy. The necessary condition for this regime to exist is 
$E_{\rm J}\ll E_{\rm C}$, which 
we assume from now on. As 
in the case of Berry's phase, the second term in eq.\ (\ref{4}) 
does not vanish because the integration contour encloses the 
singularity where the energies of several charge states 
coincide. This degeneracy occurs when $q_k=1/N$ for all $k$. 
For such $q$, the array dynamics reduces to that of a particle 
on the $N$ sites with equal energies forming a loop. The $N$ 
eigenstates of such a particle are plane waves with energies 
$\varepsilon_k=-E_{\rm J} \cos [(\varphi-2\pi k)/N]$, 
$k=0,1,\ldots,N-1$, and the equilibrium supercurrent $I(\varphi)$ 
through the array is:  
\begin{equation}
I(\varphi)=\frac{I_{\rm c}}{N} \sin \frac{\varphi}{N} \,, \;\;\;\; 
\varphi \in [-\pi,\pi]\, .
\label{7} \end{equation}   
Here $I_{\rm c}=2eE_{\rm J}/\hbar$ is the critical current of one 
junction. 
Relation (\ref{7}) should be continued periodically in 
$\varphi$ beyond the interval $[-\pi,\pi]$, and $I(\varphi)$ 
exhibits cusps at $\varphi=\pm \pi$. It also shows 
that the supercurrent decreases only as 
$N^{-2}$ at large $N$. 

Large supercurrent (\ref{7}) at resonant $q$ means that the
trajectory in $q$-space for accurate pumping should 
circle the degeneracy point sufficiently far away from it. 
It then successively brings in resonance the pairs 
of states that correspond to a Cooper pair occupying two 
neighboring islands of the array (as illustrated in Fig.\ 2 (a)
for $N=3$). If this process is slow, the Cooper 
pair is transported adiabatically between the islands by the  
usual two-state level-crossing transitions that shift it 
along the array following the gate voltages. One Cooper pair is then 
transported through the array per cycle corresponding to a $q$-space 
trajectory  
circling once around the degeneracy point. One condition  
necessary for accurate pumping is that the probability of the 
Landau-Zener transitions to the excited states is negligible 
and the array remains in the minimum-energy state throughout 
the cycle. This condition limits the rate of pumping, 
$1/\tau$, by the Josephson coupling energy, $\hbar/\tau \ll 
E_{\rm J}$. However, even then, i.e. in the regime of the present 
work, the pumping is not 
accurate due to the nonvanishing $E_{\rm J}/E_{\rm C}$. 

The gate voltages in an $N$-pump are typically \cite{b1,b3} 
sequences of triangular 
pulses shown in Fig.\ 1 (b). The array is then completely 
translationally-invariant. The pumping cycle with 
triangular gates can 
be divided into $N$ steps with Cooper pair 
transported through one junction at each step. Instead 
of calculating the charge pumped in one junction during the 
whole cycle, $Q_{\rm P}$, we can equivalently sum up the charges $Q_j$ 
transferred in all junctions during one step of pumping: 
\begin{equation}
Q_{\rm P} \equiv 2 \mbox{Re} \oint \langle m|Q_k|d m \rangle 
= \Sigma_{j=1}^{N} Q_j \, . 
\label{8} \end{equation}   
Introducing variations $\delta \langle n_j \rangle $ of the 
average number of Cooper pairs on each island during one 
step, we can express the charge conservation (which 
follows from the Hamiltonian (\ref{1}) as  $\delta \langle 
n_j \rangle =Q_j-Q_{j+1}$. Translational invariance implies 
that the distribution $\langle n_j \rangle $ is just shifted 
by one island during the step, $\delta \langle n_j 
\rangle= \langle n_{j-1} \rangle - \langle n_j \rangle $. 
For triangular gates we have precisely one Cooper pair in the 
array, $ \Sigma_{j=1}^{N} \langle n_j \rangle =1$, and we then 
obtain: $\Sigma_{j=1}^{N} Q_j = 1+N(Q_N-\langle n_N 
\rangle )$. Here the index $N$ and the adjacent 
junction are arbitrary. If we chose $N$ such 
that the island is furthest away from the junction where 
the Cooper pair is tunneling, we can neglect the 
occupation of the island in this equation and obtain:  
\begin{equation}
Q_{\rm P}=1+N Q_N \, . 
\label{10} \end{equation}   

Now we can use eqs.\ (\ref{4}) and (\ref{5}) to calculate the 
charge $Q_N$ transferred during one pumping step through the 
junction that is most distant from the tunneling Cooper pair. 
Because of the energy difference in the denominator of eq.\ 
(\ref{5}), the main contribution to $Q_{\rm P}$ arises at 
the resonances when a Cooper pair is transferred 
between the two islands. In this situation, there are two 
lowest-energy states with energy separation on the order of 
$E_{\rm J}$, and we can keep only the matrix elements of current 
between these two states in eq.\ (\ref{5}). At the two islands, 
where the Cooper pair is transferred, the resonant states are 
given by the usual expressions of the two state systems. To get 
a non-vanishing matrix element of the current in the $N$th 
junction, we must ``extend'' the wavefunctions of the resonant 
states from the two islands occupied by the Cooper pair to 
this junction. We obtain by perturbation theory in $E_{\rm J}$ 
and by eqs. (\ref{4}) and (\ref{5}): 
\begin{equation}
Q_{\rm P} = 1 - \frac{N^{N-1}(N-1)}{(N-2)!}\left(\frac{E_{\rm J}}
{2E_{\rm C}}\right)^{N-2} \!\!\!\!\!\!
\cos \varphi \, .
\label{11} \end{equation}
Thus the probability of Cooper pair tunneling limiting 
the pumping accuracy decreases roughly as $(E_{\rm J}/E_{\rm C})^{N-2}$ 
with increasing $N$. It is physically clear that this conclusion 
should remain valid for non-uniform arrays also. Results of both 
numerical (from eqs. (\ref{4}) 
and (\ref{5})) and perturbative (eq. (\ref{11})) calculations are 
shown in Table I.

\begin{figure}[ht]
\center
\includegraphics*[width=61mm]{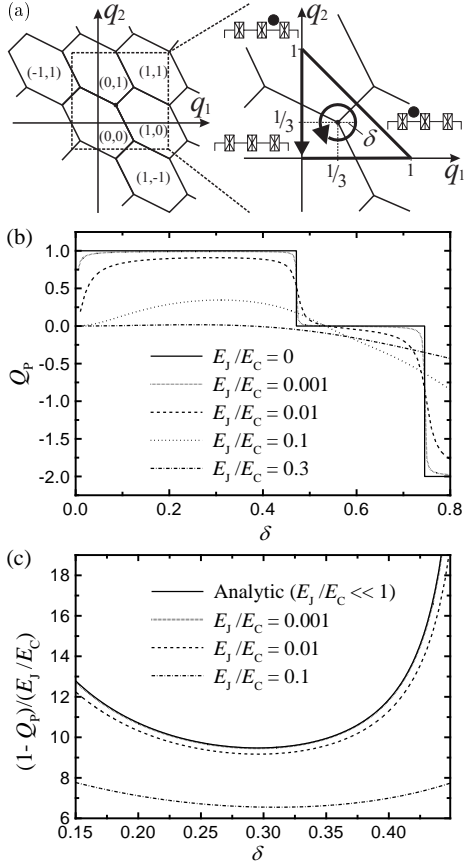}\\
\caption{\small {\bf (a)} The states with minimum charging
energy of the uniform $N=3$ pump on the $(q_1,q_2)$ plane. The 
vector $(n_1,n_2)$ denotes the stable configuration inside each 
hexagon. A circular path with radius $\delta$, and the triangular 
path of Fig.\ 1 (b) are shown. Only states contributing to the inaccuracy 
in charge transport in the leading order have been shown. {\bf (b)} and 
{\bf (c)} Numerical results of quantum inaccuracy of a 
uniform 3-pump for different values of $E_{\rm J}/E_{\rm C}$. The 
analytical result of eq.\ (\ref{13}), exact in the limit of small 
$E_{\rm J}/E_{\rm C}$, is shown by a solid line in (c).}
\end{figure}

\begin{table}[ht]
\caption{\small
Comparison between the numerically (top) and perturbatively (bottom) 
obtained transport inaccuracies ($\varphi = 0$) for uniform Cooper pair 
pumps with $N=3$, $5$, or $7$ junctions and few values of $E_{\rm J}/ 
E_{\rm C}$.} 
\label{table1}
\center
\begin{tabular}{cccc}
$E_{\rm J}/E_{\rm C}$&$N=3$&$N=5$&$N=7$\\
\hline
$0.01$&$0.0872$&$5.20\cdot 10^{-5}$&$$\\
&$0.0900$&$5.21\cdot 10^{-5}$&$1.84\cdot 10^{-8}$\\
\hline
$0.03$&$0.245$&$1.39\cdot10^{-3}$&$4.40\cdot 10^{-6}$\\
&$0.270$&$1.41\cdot 10^{-3}$&$4.47\cdot 10^{-6}$\\
\hline
$0.1$&$0.634$&$4.63\cdot 10^{-2}$&$1.60\cdot 10^{-3}$\\
&$0.900$&$5.21\cdot 10^{-2}$&$1.84\cdot 10^{-3}$\\
\hline
\end{tabular}
\end{table}

For $N=3$, triangular gate voltages correspond 
to the triangular pumping trajectory on the $(q_1,q_2)$ plane 
shown in Fig.\ 2. Another pumping method in the $N=3$ pump   
\cite{b1,b8} uses harmonic gate 
voltages, corresponding to a circular trajectory around the
degeneracy point $q_1=q_2=1/3$. It is possible to calculate the pumped 
charge directly from eqs.\ (\ref{4}) and (\ref{5}). For 
$\varphi=0$ we obtain:  
\begin{equation} 
Q_{\rm P}\!=\!1\!-\frac{3}{2}\left(\!\frac{1}{3\sqrt{2}\delta }+\! 
\frac{1}{2\!-\!3\sqrt{2}\delta }+\!\frac{1}{\frac{3}{\sqrt{5}}\delta } 
+\!\frac{1}{1\!-\!\frac{3}{\sqrt{5}}\delta }\!\right)\! \frac{E_{\rm 
J}}{E_{\rm C}},
\label{13} \end{equation}
where $\delta \equiv \left[(q_1-1/3)^2+(q_2-1/3)^2\right ]^{1/2}$ 
is the radius of the trajectory. The results of eqs.\ (\ref{13}) and 
(\ref{11}) for $N=3$ almost coincide for the optimum radius of 
$\delta \simeq 0.3$. It should be noted that the quantum inaccuracy 
in pumping is 
very significant: it is more than 20 \% at $E_{\rm J}/E_{\rm C}=0.03$, 
which 
is a very small value. (Practically, $E_{\rm J}$ is limited from below by 
temperature, while maximum $E_{\rm C}$ is limited by minimum feature size 
in fabrication.) The accurate coherent 
pumping is thus practically impossible in the $N=3$ pumps. 
Figure 2 shows $Q_{\rm P}$
calculated numerically from eq.\ (4) for $\varphi =0$ (no direct 
supercurrent present) as a function of $\delta$. For small radii the 
charge is quadratic in $\delta$, $Q_{\rm P}/\pi \delta^2  = (8E_{\rm 
C}/27E_{\rm J})^2$, as can be derived from eq.\ (\ref{4}). At large 
$\delta$'s the pumped charge in Fig.\ 2 starts to decrease  
since the trajectory approaches another degeneracy point at $q_1=q_2= 
2/3$. 

Finally, we consider supercurrent in the regime of accurate pumping. In 
this regime, current is largest during the Cooper pair transition 
between a pair of islands of the array. The effective tunneling 
amplitude of this transition consists of two parts: the 
direct transition between the islands through one junction 
separating them, and the transition through the rest 
$N-1$ junctions. Interference between these two 
processes determines the phase-dependent part of the 
Cooper pair energy. We obtain the supercurrent of $(N-1)$-junction 
tunneling by standard perturbation theory: 
\begin{equation} 
I(\varphi)=I_{\rm c} \frac{E_{\rm J}}{(\varepsilon^2+E_{\rm J}^2)^{1/2}} 
\frac{N^{N-2}(N\!\!-\!1)}{2(N\!\!-\!2)!}\!\left(\frac{E_{\rm J}}{2E_{\rm 
C}}\right)^{N-2}\!\!\!\!\!\!\! \sin 
\varphi , 
\label{15} \end{equation}
where $\varepsilon$ is the energy difference between the Cooper 
pair states on the two islands of resonant transition. Current (\ref{15}) 
at resonance scales as 
$(E_{\rm J}/E_{\rm C})^{N-2}$, but since the transition region represents 
only a fraction of the pumping cycle on the order of $E_{\rm J}/E_{\rm 
C}$, 
the overall supercurrent contribution to the pumped charge 
(\ref{4}) scales as $(E_{\rm J}/E_{\rm C})^{N-1}$. This means that the 
inaccuracy is dominated by quantum tunneling of Cooper pairs (given by 
eqs. (\ref{11}) or 
(\ref{13})) even with nonzero $\varphi$ if the pumping is not too slow. 

In conclusion, we have developed a quantitative theory of  
adiabatic Cooper pair transport in one-dimensional arrays of 
Josephson junctions. The theory predicts, among other things, 
that the quantum inaccuracy of the Cooper pair pumping in arrays 
with a small number of junctions is very large, the fact that can 
explain lack of success of the experimental attempt \cite{b8} to 
pump Cooper pairs.

This work was supported by AFOSR of the U.S. and the Academy of Finland.

\end{document}